\documentstyle[twoside,fleqn,epsf,espcrc2]{article}

\def\be         {\begin{equation} }
\def\ee         {\end{equation} }
\def\bea                {\begin{eqnarray} }
\def\eea                {\end{eqnarray} }

\def\Tr                 {\mathop{\rm Tr}}

\newlength{\figsize}
\newlength{\figoffset}
\newlength{\figbackup}
\newlength{\figendsp}

\def\preprints{
\vspace{-10ex}
{\small
\begin{tabbing}
\` Edinburgh 99/11 \\
\` September 1999 \\
\end{tabbing} 
}
\vspace*{0.1in}
}

\title{
\preprints
The topological susceptibility in `full' (UK)QCD.}

\author{\underline{A. Hart}%
        \address{Dept. of Physics, University of Edinburgh, 
          King's Buildings, Edinburgh EH9 3JZ, Scotland.}
        and 
      M. Teper%
        \address{Theoretical Physics, University of Oxford, 
        1 Keble Road,
        Oxford OX1 3NP, England.\\}.}

\begin{document}

\begin{abstract}
\noindent
We report first calculations of the topological susceptibility
measured using the field theoretic method on SU(3) gauge
configurations produced by the UKQCD collaboration with two flavours
of dynamical, improved, Wilson fermions. Using three ensembles with
matched lattice spacing but differing sea quark mass we find that
hybrid Monte Carlo simulation appears to explore the topological
sectors efficiently, and a topological susceptibility consistent with
increasing linearly with the quark mass.
\end{abstract}

\maketitle

The latest UKQCD runs generate SU(3) field configurations using a
Wilson gauge action coupled to $n_f = 2$ flavours of Wilson sea
quarks, non--perturbatively improved such that the leading order
discretisation errors in spectral quantities should be ${\cal
O}(a^2)$. In dynamical simulations the lattice spacing, $a$, is
influenced by the gauge coupling, $\beta$, and the quark mass
parameter, $\kappa$. The ensembles studied represent three points on a
trajectory in the $(\beta,\kappa)$ space of (approximately) constant
$a$ and physical volume ($V a^4 = 16^3.32 a^4$)%
~\cite{irving_matching}. 
[The lattice spacing has been defined from the static inter--quark
potential, using $r_0 = 0.49$~fm%
~\cite{sommer_scale}.]
Along such trajectories, however, there is a change in chiral
behaviour as seen in the pseudoscalar to vector mesonic
mass ratio in Table~\ref{tab_params}. [Dimensionless lattice quantities
are denoted with circumflexes throughout.]

We measure the topological charge density on the lattice using the field
theoretic operator
\be
\hat{Q}(n) = \frac{1}{2} \times \frac{1}{16} 
\varepsilon^\pm_{\mu \nu \sigma \tau} 
\Tr U_{\mu \nu}(n) U_{\sigma \tau}(n),
\ee
symmetrised such that 
$
\varepsilon^\pm_{\mu \nu \sigma, -\tau} = 
- \varepsilon^\pm_{\mu \nu \sigma, +\tau}
$
{\it etc.} to improve the signal on moderately cooled lattices. The
topological charge, $\hat{Q}$, and susceptibility, $\hat{\chi}$, are
then
\be
\hat{Q} = \frac{1}{32 \pi^2} \sum_n \hat{Q}(n)
\mbox{\hspace{2em}}
\hat{\chi} = \frac{\langle \hat{Q}^2 \rangle}{V}. 
\ee
\begin{table}[b]

\vspace{\figendsp}
\begin{tabular}{llll}
\hline \hline
$\beta$ &       5.20 &          5.26 &          5.29 \\
$\kappa$ &      0.1350 &        0.1345 &        0.1340 \\
No. cfgs. &     150 &           100 &           100 \\
$\hat{r}_0$ &   4.58 (8) &      4.58 (6) &      4.45 (6) \\
$\frac{\hat{m}_\pi}{\hat{m}_\rho}$ &
                0.69 (1) &      0.79 (1) &      0.83 (1) \\
\hline \hline
\end{tabular}
\caption{Lattice spacing and pion to rho mass ratios
\cite{garden_lat99}.
}
\label{tab_params}
\end{table}

Although formally in the weak coupling limit $\hat{Q}$ is related
to that of the continuum by
$
\hat{Q} = a^4 Q + {\cal O}(a^6),
$
the prefactors of the higher order terms typically have momenta on the
scale of $1/a$ such that they are ${\cal O}(1)$, and $\hat{Q}$ becomes
increasingly dominated by ultraviolet noise near the continuum limit.
It is also subject to a multiplicative renormalisation that reduces
the signal at the $\beta$ couplings accessible to simulation.

An established, and well understood, procedure to extract the
topological signal is to cool the configurations prior to measuring
$\hat{Q}$. This locally smoothens the lattice fields to remove the
ultraviolet fluctuations, and drives the renormalisation constant to
unity. We move through the lattice links in a staggered fashion,
updating each Cabibbo--Marinari subgroup in turn so as to exactly
minimise the SU(2) Wilson plaquette action. This action is the most
local and thus is expected to do least damage to correlation functions
on physical length scales. Updating every link once corresponds to
one cooling `sweep'.

Such a cooling action, however, also destroys topological features
when applied {\it in extremis}. Instanton--anti-instanton pairs in the
vacuum are not stable minima of the action (even in the continuum) and
under cooling there is an attractive force leading to
annihilation. Whilst this is an issue in measurements of instanton
size distributions, there in no net change in the topological charge
and the susceptibility is thus stable.

\begin{figure}[t]

\leavevmode
\begin{center}

\hbox{%
\hspace{\figoffset}
\epsfxsize = \figsize
\epsffile{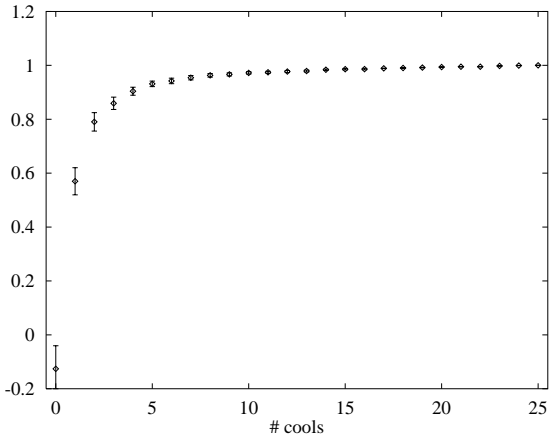}
}

\vspace{\figbackup}
\end{center}

\caption{Normalised $\langle \hat{Q}(n_c)\cdot\hat{Q}(25) \rangle$
versus the number of cools, $n_c$, at $\beta=5.20$.}
\label{fig_qcorrs}

\vspace{\figendsp}
\end{figure}
\begin{figure}[t]

\leavevmode
\begin{center}

\hbox{%
\hspace{\figoffset}
\epsfxsize = \figsize
\epsffile{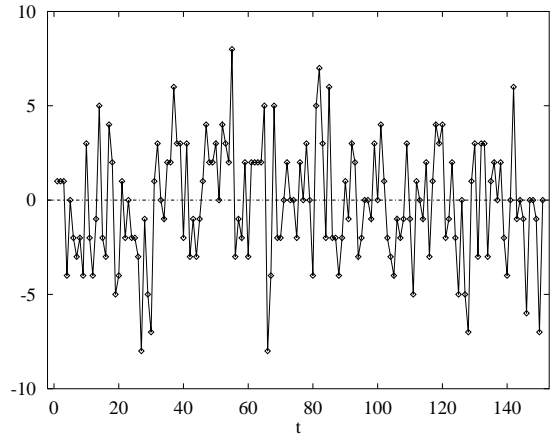}
}

\vspace{\figbackup}
\end{center}

\caption{$\hat{Q}(n_c=10)$ vs. simulation time in
units of 40 HMC trajectories at $\beta=5.20$.}
\label{fig_thist}

\vspace{\figendsp}
\end{figure}

The lattice regularisation breaks scale invariance, and the Wilson
action of an instanton is an increasing function of the core size,
$\rho$. (Isolated) instantons shrink and disappear under cooling,
leading to a net change in $\hat{Q}$ and $\hat{\chi}$. Whilst such
events may be monitored in relatively smooth configurations, it is
still desirable to perform as few cooling steps as possible to expose
the signal before making measurements. In Fig.~\ref{fig_qcorrs} we
plot the normalised correlation between the topological charge after a
given number of cooling sweeps, $n_c$, relative to that after
$n_c=25$:
\be
\frac{\langle \hat{Q}(n_c)\cdot\hat{Q}(25) \rangle}
{\frac{1}{2} \left(\langle \hat{Q}(n_c) \rangle^2 +
\langle \hat{Q}(25) \rangle^2 \right) }
\ee
We find remarkable stability in $\hat{Q}$ from 5 cooling
sweeps out to at least 25 cooling steps.

One technical point of interest is the rate at which configurations
become topologically independent under Monte Carlo updating; the
topological charge is related to the small eigenvalues of the fermion
matrix and should be one of the slowest modes to decorrelate. We show
a Monte Carlo time history plot for our most chiral ensemble in
Fig.~\ref{fig_thist}, and we estimate the integrated autocorrelation
times in units of the 40 HMC trajectories between configurations in
Table~\ref{tab_susc}, albeit using a small ensemble, indicating pretty
good decorrelation at these quark masses. We note that autocorrelation
times are longer for the larger $\beta$ ensembles despite these being
the less chiral.

\begin{figure}[t]

\leavevmode
\begin{center}

\hbox{%
\hspace{\figoffset}
\epsfxsize = \figsize
\epsffile{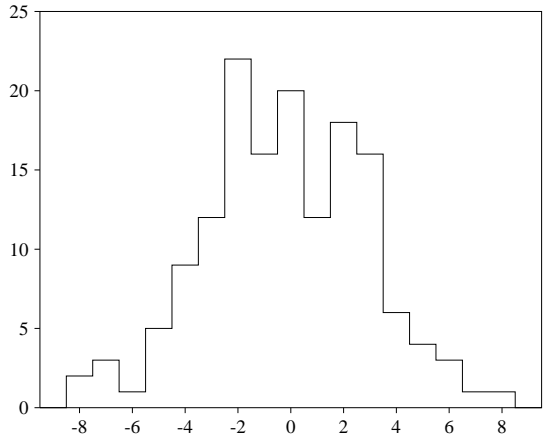}
}

\vspace{\figbackup}
\end{center}

\caption{Histogram of $\hat{Q}^{\rm int}$ at 
$n_c=10$ on the $\beta = 5.20$ ensemble.}
\label{fig_hist}

\vspace{\figendsp}
\end{figure}

Also of issue is the ergodicity of the MC update. Measuring $\hat{Q}$
over the whole configuration we find the ensemble averages to be
within $\sim 1.5$ (relatively large) standard deviations of zero in
Table~\ref{tab_susc}, and Fig.~\ref{fig_hist} shows a relatively
Gaussian sampling of the topological sectors.

In the continuum, the topological charge of a configuration is
integer. On the lattice this is not so, nor is the charge measured
using our operator on a configuration particularly close to integer.
This can be attributed to the presence of narrow instantons whose
charge is significantly less than unity. Attempts can be made to
calibrate and correct for this
\cite{smith_teper}
but we defer application of such procedures until
\cite{hart_progress}.
We have both maintained the charge as a non--integer for the purposes
of calculating the topological susceptibility, and also rounded the
value on each configuration to the nearest integer, and found that the
topological susceptibilities obtained are consistent within
statistical errors.

\begin{table}[b]

\vspace{\figendsp}
\begin{tabular}{llrr}
\hline \hline
\multicolumn{1}{c}{$\beta$} &       
\multicolumn{1}{c}{$\tau_{\rm int}^{\rm est.}$} &       
\multicolumn{1}{c}{$\langle \hat{Q} \rangle$} &     
\multicolumn{1}{c}{$\hat{\chi}^{\rm int} \cdot 10^{5}$} \\
\hline
5.20 &          0.71 (41) &     0.03 (56) &             7.5 (1.4) \\
5.26 &          0.67 (28) &     $-0.47$ (28) &          7.2 (1.5) \\
5.29 &          1.5 (1.2) &     1.10 (97) &             14.8 (2.3) \\
\hline \hline
\end{tabular}

\caption{Mean $\hat{Q}$, autocorrelation time and the
topological susceptibility at $n_c = 25$.}
\label{tab_susc}
\end{table}

\begin{figure}[t]

\leavevmode
\begin{center}

\hbox{%
\hspace{\figoffset}
\epsfxsize = \figsize
\epsffile{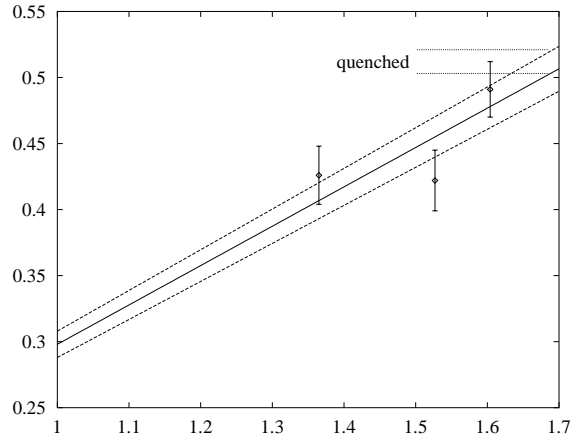}
}

\vspace{\figbackup}
\end{center}

\caption{$\hat{r}_0 (\hat{\chi}^{\rm int})^{\frac{1}{4}}$ vs. 
$\sqrt {\hat{r}_0 \hat{m}_\pi}$ with a linear fit.}
\label{fig_susc}

\vspace{\figendsp}
\end{figure}

In Table~\ref{tab_susc} we show our estimates for the topological
susceptibility measured after 25 cooling sweeps. In the chirally
broken, confining phase at low temperatures, the sea quarks induce an
attractive interaction which leads to instanton--anti-instanton pairing
and a suppression of the topological susceptibility as the dynamical
quark mass is lowered:
\be
\chi = \frac{m_q \langle \bar{\psi}\psi \rangle}{n_f^2} +
{\cal O}(m_q^2)
\ee
where $\langle \bar{\psi}\psi \rangle$ is summed over light quark
flavours, and should be evaluated in the $m_q \to 0$ limit.  The quark
mass is not known {\it a priori} but may be re-expressed in terms of
the pseudoscalar decay constant using the PCAC relation
$
m_q \langle \bar{\psi}\psi \rangle = f_\pi^2 m_\pi^2
$
such that for sufficiently chiral sea quarks, the topological
susceptibility should be quadratic in the pseudoscalar mass and decay
constant:
\be
\chi = \frac{f_\pi^2 m_\pi^2}{n_f^2} + {\cal O}(m_\pi^4).
\ee
In Fig.~\ref{fig_susc} we plot the susceptibility versus the
pseudoscalar mass in units of $r_0$, and find the data are consistent
with such a leading order fit through the origin. We may use the
fitted slope of this graph to provide an estimate of the pseudoscalar
decay constant, finding it to be very low compared to the experimental
value.

Although such an estimate must be regarded as preliminary given the
volume of data so far analysed, it is clear from Fig.~\ref{fig_susc}
that the heaviest of our sea quark masses yields a topological
susceptibility that is in statistical agreement with the quenched
value, and attempting to fit the leading order chiral behaviour to
this point is likely to lead to an underestimate of $f_\pi$; ${\cal
  O}(m_q^2)$ terms are likely to be large, including the effects of
not extrapolating $\langle \bar{\psi} \psi \rangle$ to $m_q \to 0$.
Ongoing analysis of further configurations at these and lighter sea
quark masses should clarify the situation.

\newpage
\vspace{5ex}
\noindent
{\bf Acknowledgments.}

\vspace{2ex}
\noindent
A.H.'s work was supported in part by United Kingdom PPARC grant
GR/K22744, and that of M.T. by grants GR/K55752 and GR/K95338.

\end{document}